\documentclass{article}
\usepackage[utf8]{inputenc}

\title{Thermodynamic Behavior of Time-Dependent Wormholes in Palatini $f(R)$ Gravity}

\author{Hamidreza Saiedi}

\date{}

\usepackage{natbib}
\usepackage{graphicx}

\begin{document}

\maketitle

Department of Physics, Florida Atlantic University,  Boca Raton, FL, USA 

hsaiedi2014@fau.edu  ;  hrssaiedi@gmail.com

\vspace{30pt}

\begin{abstract}
In the frame work of Palatini approach to modified $f(R)$ theories of gravity, we attempt to investigate
the thermodynamic behavior of time-dependent wormhole geometries. By considering an evolving wormhole
spacetime, we find the field equations in Palatini $f(R)$ gravity. The energy density and pressure are
redefined in order to satisfy the continuity equation. To discuss the thermodynamic behavior, an expression
for the variation of total entropy is obtained. Then, we extend our discussions to apparent and event
horizons of evolving wormholes. We employ the radius of apparent and event horizons to determine
the validity of generalized second law (GSL) of thermodynamics which states that the variation of  total entropy is non-negative.

\end{abstract}  
\section{Introduction}
The Morris-Thorne (MT) wormholes have been vastly considered in many recent research projects[1]. The MT wormholes are limited to static and spherically  symmetric spacetimes. In general, according to General Relativity (GR), these wormholes need an exotic matter (negative-energy matter) which violates the weak and null energy conditions in order to be traversable. It means that, without introducing an exotic matter, these wormholes are unstable and travel in and out of the  throat would be impossible. A class of required exotic matter to support traversable wormholes could be phantom energy which is investigated in[2-4].
The wormhole solutions and their energy conditions have been extensively studied in different theories of gravity without need of an exotic matter[5-15]. Also, traversability of time-dependent wormholes in the absence of negative-energy matter has been carefully investigated in modified $f(R)$ gravity (metric approach)[16].
However, in metric formalism of $f(R)$ gravity the metric is only considered as
dynamical variable and the action is varied respect to the metric. In Palatini approach, the
connection $\Gamma^\lambda_{\mu\nu}$ and the metric $g_{\mu\nu}$ are dynamical variables and
the action is varied respect to them separately. 
\vspace{10pt}

The discovery of connection between thermodynamics and gravity has been well established in theoretical physics.
For instance, the validity of thermodynamics laws and thermodynamics of dark energy have been the interesting
subjects of numerous studies in the recent and past years[17-24]. Regarding the second law, which is extended
to ``generalized second law of thermodynamic" in black holes, stating that the sum of black hole horizon
entropy and the usual entropy is a non-decreasing function, one can employ it to the universe or wormhole horizons[25-27].
It has been shown  that for cosmological horizons the thermodynamics laws are valid.
Generally, the validity of this statement has been proven in Einstein theory of gravity. In the case of
modified gravitational theories, one can apply it to find out constraints. Authors in [28] have investigated
the generalized second law of thermodynamics in Einstein general relativity. This attempt has been also made
in Scalar-Gauss-Bonnet gravity [29] and Lancos-Lovelock gravity [30] , and as a result, a clear formalism to
understand the horizon thermodynamics in spherically symmetric spacetimes was expanded.
In literature, for wormhole spacetimes, thermodynamic behavior of horizons has been discussed in Einstein
gravitational field equations. In the context of modified $f(R)$ theory, author in [31] investigated the wormhole thermodynamics
at the apparent horizon.  In this frame work, Y. Zheng and R. Yang  have studied the horizon thermodynamics
for black hole geometry[32]. The thermodynamics of an evolving lorentzian wormhole with entropy correction
has been also discussed in[33]. Additionally, in $f(T,T_G)$ gravity, the generalized second law of thermodynamics
has been investigated by Zubair and Jawad[34]. 
\vspace{10pt}

All these explanations prove the deep connection not only
between thermodynamics and Einstein gravity theory, but also between thermodynamics and various theories of gravity.
Therefore, in this work we aim to investigate this relation in Palatini version of
$f(R)$ gravity for time-dependent wormhole spacetimes and obtain an expression for the variation of
total entropy at wormhole horizons. Then, the generalized second law of thermodynamics has been well
discussed at the apparent and event horizons of evolving wormholes. 
\vspace{30pt}

\section{Evolving  Wormholes in Palatini Formalism}
Wormholes consist of tunnels in spacetime with nontrivial topology that link two or more regions of
the same or different spacetimes. They have two mouths which are coincided as the throat of
a Morris-Thorne (MT) wormhole  in the static case. The MT wormholes require an exotic fluid violating
the null energy condition in order to keep the wormhole's throat open and hold up the wormhole geometry.
A simple generalized MT wormhole spacetime to the dynamical background is written as[35] 
\begin{equation}
ds^2 = - e^{2\Phi(t,r)} dt^2 + a^2(t) \left[ \frac{dr^2}{1 - \frac{b(r)}{r}
} +r^2 d\Omega^2_2  \right] \ ,
\end{equation}
where $d\Omega^2_2 = d\theta^2 +  \sin^2 \theta d\phi^2$. $a(t)$ is the scale factor, $ \Phi(t,r)$ and  $b(r)$ are
called the potential and shape functions of the wormhole, respectively. It can be seen from the above metric that
when $ \Phi(t,r) \rightarrow  \Phi(r)$ and $a(t) \rightarrow constant$ it represents the static MT wormhole. Furthermore,
if   $b(r) \rightarrow Kr^3$ and $ \Phi(t,r) \rightarrow  constant$, it turns out to the Friedmann-Robertson-Walker (FRW) metric.
For simplicity, we assume that $ \Phi(t,r)=0$ in the rest of the paper.
In modified $f(R)$ theories of gravity, the action has the form 
\begin{equation}
A = \frac{1}{16\pi G}\int {\sqrt{-g} \ f(R) \ d^4x} \ + A_m[g_{\mu\nu}, \psi_m]
\end{equation}
$A_m[g_{\mu\nu}, \psi_m]$ is the matter action. $g_{\mu\nu}$ and   $\psi_m$ are the metric and the matter field, respectively.
In Palatini $f(R)$ gravity, the metric $g_{\mu\nu}$ and the connection $\Gamma^\lambda_{\mu\nu}$  are considered
as dynamical variables to be varied independently. Varying the action (2) respect to the connection leads to 
\begin{equation}
\nabla_\alpha(\sqrt{-g} F g^{\mu\nu}) = 0      \ ,
\end{equation}
and  respect to the metric  yields  
\begin{equation}
FR_{\mu\nu}(\Gamma) - \frac{1}{2} f(R) g_{\mu\nu} = 8\pi G T_{\mu\nu}^{(m)} \ ,
\end{equation} 
where $R_{\mu\nu}(\Gamma)$ represents the Ricci tensor corresponding to the connection $\Gamma^\lambda_{\mu\nu}$, $T_{\mu\nu}^{(m)}$ is
the usual energy-momentum tensor, and $F=\frac{df}{dR}$. We consider $k^2 = 8\pi G = 1$ just for simplicity.
Under conformal transformations of the Ricci tensor, equations (3) and (4) lead to the following relation [36-38]. 

$$ R_{\mu\nu}(g) - \frac{1}{2} g_{\mu\nu} R(g) = \frac{1}{F} T_{\mu\nu}^{(m)} - \frac{R(T)F - f }{2F} g_{\mu\nu} + \frac{1}{F}\left (\nabla_\mu\nabla_\nu F - g_{\mu\nu}\nabla_\alpha\nabla^\alpha F\right ) $$

\begin{equation}
+  \frac{3}{2(F)^2} \left [ - \partial\mu F \partial\nu F  +   \frac{1}{2} g_{\mu\nu} (\partial F)^2    \right ]
\end{equation}

$R_{\mu\nu}(g)$ and $R(g) = g^{\mu\nu}R_{\mu\nu}(g)$ are the usual Ricci tensor and scalar curvature,
and  $R(T) = g^{\mu\nu}R_{\mu\nu}(\Gamma)$ is given by 
\begin{equation}
R(T) = R(g) + \frac{3}{2(F)^2} \partial_\lambda F \partial^\lambda F - \frac{3}{F} \nabla_\alpha\nabla^\alpha F \ \ .
\end{equation}
After some simplifications and using the above relation, equation (5) can be written as
\begin{equation}
G_{\mu\nu}= R_{\mu\nu}(g) - \frac{1}{2} g_{\mu\nu} R(g) =  \frac{T_{\mu\nu}^{(m)}}{F} + T_{\mu\nu}^c  \ .
\end{equation}
where 
$$ T_{\mu\nu}^c = \frac{1}{F} \left [\nabla_\mu\nabla_\nu F - \frac{3}{2F}(\partial_\mu F \partial_\nu F) \right ] $$
\begin{equation}
- \frac{1}{F} \left [ \frac{1}{4} g_{\mu\nu} \left ( T + FR_g +  \nabla_\alpha\nabla^\alpha F  - \frac{3}{2F}(\partial F)^2     \right)  \right ] ,
\end{equation}

\vspace{10pt}

and $T_{\mu\nu}^m$ is the energy-momentum tensor with $\rho(t,r)$(energy density), 
$p_r(t,r)$ (radial pressure), and $p_t(t,r)$(tangential pressure) which is given as 

\begin{equation}
T^{\nu (m)}_{\mu} = diag ( - \rho(t,r), p_r(t,r), p_t(t,r), p_t(t,r) ) \ ,
\end{equation} 
The field equations (7) yield the following relations for density, radial and tangential pressures. 
\begin{eqnarray}
\rho &=& \frac{1}{4}  \left ( T + FR_g +  \nabla_\alpha\nabla^\alpha F  - \frac{3}{2F}(\partial F)^2     \right) \nonumber \\ &+& 3H^2F + \frac{b'F}{a^2r^2}  -   \ddot{F} + \frac{3}{2F}\dot{F}^2  \ , \\ \nonumber \\
p_r &=& \frac{1}{4}  \left ( T + FR_g +  \nabla_\alpha\nabla^\alpha F  - \frac{3}{2F}(\partial F)^2     \right)  + \frac{3(r-b)F'^2}{2Fa^2r}      \nonumber \\ &-&  2\dot{H}F - 3H^2F - \frac{bF}{a^2r^3}   + H\dot{F} - \frac{(r-b)F''}{a^2r}  +  \frac{(b'r-b)F'}{2a^2r^2}   \ , \\ \nonumber \\
p_t &=& \frac{1}{4}  \left ( T + FR_g +  \nabla_\alpha\nabla^\alpha F  - \frac{3}{2F}(\partial F)^2     \right)  \nonumber \\  &-&  2\dot{H}F - 3H^2F - \frac{(b'rF-bF)}{2a^2r^3}  + H\dot{F} - \frac{(r-b)F'}{a^2r^2}  \ ,
\end{eqnarray}
where $F=F(r,t)$, $\dot{F}=(\partial/\partial t)F(r,t)$, $\ddot{F}=(\partial/\partial t)^2F(r,t)$, $F' = (\partial/\partial r)F(r,t)$, and $F'' = (\partial/\partial r)^2F(r,t)$.
To satisfy the continuity equation, one can reorganize the equations (10), (11), and (12) as follow.
\begin{eqnarray}
\rho &=& 3H^2F - \ddot{F} + \frac{b'F}{a^2r^2} + \frac{3}{2F}\dot{F}^2   \ , \\ \nonumber \\
p_r &=&  -  2\dot{H}F - 3H^2F + H\dot{F} - \frac{bF}{a^2r^3}    - \frac{(r-b)F''}{a^2r}  \nonumber \\ &+&  \frac{(b'r-b)F'}{2a^2r^2} + \frac{3(r-b)F'^2}{2Fa^2r}  \ , \\ \nonumber \\
p_t &=&  -  2\dot{H}F - 3H^2F + H\dot{F} - \frac{(b'rF-bF)}{2a^2r^3}   - \frac{(r-b)F'}{a^2r^2}  \ ,
\end{eqnarray} 
\vspace{20pt}

\section{Thermodynamics of Evolving Wormholes}
The similarities between thermodynamics and black hole physics urge us to consider more general
spacetimes like evolving wormholes. It is shown that thermodynamics laws can be applicable to black holes
horizons by introducing new quantities such as entropy and surface gravity. In these kinds of studies, the
surface gravity is proportional to a temperature and the entropy corresponds to the horizon area
of the black holes. In the following, we employ non-static wormhole geometry and discuss its thermodynamic
behavior at the apparent and event horizons in Palatini $f(R)$ gravity theories. It will be shown that
generalized second law of thermodynamics at these horizons will be valid. To study the thermodynamic
properties of a time-dependent wormhole, let us consider the metric (1) as 
\begin{equation}
ds^2 = h_{ij} dx^i dx^j + \tilde{r}^2 d\Omega^2_2  \ \ \ \ , \ \ \ \ (i,j = 0, 1)
\end{equation} 
where  $   \tilde{r}=a(t)r$ represents the physical radius and  
\begin{equation}
h_{ij} = diag  \left[-1 ,  a^2(t) \left( 1 - \frac{b(r)}{r}  \right)^{-1} \right] \ .
\end{equation}
Now, we consider the first law of thermodynamics on the horizon defined as 
\begin{equation}
T_hdS_h = dQ = -dE_h
\end{equation} 
where $S_h$ and  $T_h$ are the entropy of the horizon and its temperature, respectively.
$dQ$ is the heat flow through the horizon and $dE_h$ represents the amount of energy crossing the horizon.
The unified first law says that[39] 
\begin{equation}
dE = A\Psi + WdV
\end{equation} 
where  $V=\frac{4}{3}\pi\tilde{r}^3$ is the volume and  $A=4\pi\tilde{r}^2$ is the area.
$\Psi = \Psi_i dx^i$ is defined as the energy flux generated by the surrounding material with the
energy-supply vector $\Psi_i$ given as 
\begin{equation}
\Psi_i = T_i^j\partial_j\tilde{r} + W \partial_i\tilde{r}
\end{equation} 
and $W$ is the work density defined as 
\begin{equation}
W = - \frac{1}{2} h^{ij}T_{ij} = \frac{1}{2} (\rho - p_r)
\end{equation}
So, energy flux $\Psi$ can be obtained as 
\begin{equation}
\Psi = \Psi_i dx^i = \frac{1}{2} ( \rho + p_r )d\tilde{r} - \tilde{r}H ( \rho + p_r )dt
\end{equation} 
Substituting (22) and (23) into (20), we have the following expression for the energy crossing on the horizon. 
\begin{equation}
dE_h = 4\pi\tilde{r}_h^2  \rho d\tilde{r}_h - 4\pi\tilde{r}_h^3H ( \rho + p_r )dt
\end{equation}
Therefore, the entropy variation on the horizon follows as 
\begin{equation}
T_hdS_h =  4\pi\tilde{r}_h^3H ( \rho + p_r )dt - 4\pi\tilde{r}_h^2  \rho d\tilde{r}_h
\end{equation} 
By inserting $\rho(t,r)$ and $p_r(t,r)$ from (13) and (14), the above equation becomes
\begin{eqnarray}
T_h\dot{S}_h &=& 4\pi\tilde{r}_h^3H \left[     - \ddot{F} + \frac{ab'F}{\tilde{r}_h^2} + \frac{3}{2F}\dot{F}^2  -  2\dot{H}F  + H\dot{F} - \frac{abF}{\tilde{r}_h^3}   \right] \ \nonumber \\ \nonumber \\ &+& 4\pi\tilde{r}_h^3H\left[- \frac{(\tilde{r}_h-ab)F''}{\tilde{r}_h}  +  \frac{(ab'\tilde{r}_h-ab)F'}{2\tilde{r}_h^2} + \frac{3(\tilde{r}_h-ab)F'^2}{2F\tilde{r}_h}      \right]  \ \nonumber \\ \nonumber \\ &-&  4\pi\tilde{r}_h^2 \left[   3H^2F - \ddot{F} + \frac{ab'F}{\tilde{r}_h^2} + \frac{3}{2F}\dot{F}^2    \right] \dot{\tilde{r}}_h
\end{eqnarray}
Here $b = b(\frac{\tilde{r}}{a})$,  $F = F(\frac{\tilde{r}}{a},t)$, $b' = (\partial/\partial \tilde{r}) b(\frac{\tilde{r}}{a})$  , $\dot{F}=(\partial/\partial t)F$, $\ddot{F}=(\partial/\partial t)^2F$, $F' = (\partial/\partial \tilde{r})F$, and $F'' = (\partial/\partial \tilde{r})^2F$. \\
The Gibbs equation in thermodynamics states that[40] 
\begin{equation}
T_h dS_I =  d(\rho V) + p dV \ ,
\end{equation}
Here the entropy within the horizon denotes as $S_I$ and $p$ represents as the average pressure inside the horizon
which can de written as $p = (p_r + 2p_t)/3$.
The energy conservation equation ($T^{\mu}_{\nu ; \mu}$) says that 
\begin{equation}
\dot{\rho} + H (3\rho + p_r + 2p_t) = 0
\end{equation}
Using the above equation, one can obtain the following relation for the variation of internal entropy. 
\begin{equation}
T_h dS_I = - \frac{4\pi \tilde{r}^2_h}{3} (3\rho + p_r + 2p_t + \frac{\tilde{r}_h\rho^{'}}{a})(H\tilde{r}_h dt - d\tilde{r}_h) \ ,
\end{equation}
($\rho^{'} = \partial \rho / \partial r$). Using (13), (14) and (15), the equation (28) can be rewritten as 
\begin{eqnarray}
T_h \dot{S}_I &=&  \frac{4\pi \tilde{r}^2_h}{3}(\dot{\tilde{r}}_h - H\tilde{r}_h) \left[     - 3\ddot{F} + \frac{2ab'F}{\tilde{r}_h^2} + \frac{9}{2F}\dot{F}^2  -  6\dot{H}F  + 3H\dot{F}      \right]  \ \nonumber \\
&+& \frac{4\pi \tilde{r}^2_h}{3}(\dot{\tilde{r}}_h - H\tilde{r}_h) \left[    \frac{(ab - \tilde{r}_h)F''}{\tilde{r}_h} +  \frac{(3ab'\tilde{r}_h+3ab)F'}{2\tilde{r}_h^2} \right]  \ \nonumber \\
&+& \frac{4\pi \tilde{r}^3_h}{3}(\dot{\tilde{r}}_h - H\tilde{r}_h)  \left[  3H^2F' - (\ddot{F})' + \frac{ab''F }{\tilde{r}_h^2}   + \frac{3F(\dot{F}^2)' - 3F'\dot{F}^2}{2F^2}        \right] \ \nonumber \\
&+& \frac{4\pi \tilde{r}^2_h}{3}(\dot{\tilde{r}}_h - H\tilde{r}_h) \left[     \frac{3(\tilde{r}_h-ab)F'^2}{2F\tilde{r}_h}  - \frac{2F'}{\tilde{r}_h}         \right] 
\end{eqnarray} 
\vspace{20pt}

\subsection{Thermodynamics at Apparent Horizon}
Now, in Palatini approach to modified $f(R)$ gravity we aim to find an expression for the
variation of total entropy $T_h\dot{S}_{tot} = T_h (\dot{S}_I + \dot{S}_h)$ at the
apparent horizon which is determined by  $h^{ij} \partial_i\tilde{r} \partial_j\tilde{r} = 0 $.
So, the dynamical apparent horizon can be evaluated as 
\begin{equation}
H^2 \tilde{r}_A^2 + \frac{ab(\frac{\tilde{r}_A}{a})}{\tilde{r}_A} -1  = 0 \ ,
\end{equation}
where $\tilde{r}_A$ denotes the apparent horizon radius. By combining (25) and (29) in order to have
the variation of total entropy, we reach 
\begin{eqnarray}
T_h\dot{S}_{tot} &=&  \frac{4\pi \tilde{r}^2_h}{3}(\dot{\tilde{r}}_h - H\tilde{r}_h) \left[  -  6\dot{H}F - 9H^2F + 3H\dot{F} - \frac{3ab'\tilde{r}_hF}{\tilde{r}_h^3}         \right]  \ \nonumber \\
&+& \frac{4\pi \tilde{r}^2_h}{3}(\dot{\tilde{r}}_h - H\tilde{r}_h) \left[    \frac{(ab - \tilde{r}_h)F''}{\tilde{r}_h} +  \frac{(3ab'\tilde{r}_h+3ab-4\tilde{r}_h)F'}{2\tilde{r}_h^2}  \right]  \ \nonumber \\
&+& \frac{4\pi \tilde{r}^2_h}{3}(\dot{\tilde{r}}_h - H\tilde{r}_h) \left[    \frac{3(\tilde{r}_h-ab)F'^2}{2F\tilde{r}_h}  \right]  \ \nonumber \\
&+& \frac{4\pi \tilde{r}^3_h}{3}(\dot{\tilde{r}}_h - H\tilde{r}_h)  \left[  3H^2F' - (\ddot{F})' + \frac{ab''F }{\tilde{r}_h^2}   + \frac{3F(\dot{F}^2)' - 3F'\dot{F}^2}{2F^2}        \right] \ \nonumber \\
&+& 4\pi\tilde{r}_h^3H\left[- \frac{(\tilde{r}_h-ab)F''}{\tilde{r}_h}  +  \frac{(ab'\tilde{r}_h-ab)F'}{2\tilde{r}_h^2} + \frac{3(\tilde{r}_h-ab)F'^2}{2F\tilde{r}_h}      \right]  \ \nonumber \\ \nonumber \\
&+& 4\pi\tilde{r}_h^3H \left[   -  2\dot{H}F -3H^2F  + H\dot{F} - \frac{abF}{\tilde{r}_h^3}   \right]
\end{eqnarray} 
From (30), by taking the derivative, one can check that 
\begin{equation}
2H\dot{H}\tilde{r}_A^2 + 2H^2\tilde{r}_A\dot{\tilde{r}}_A = \frac{ab\dot{\tilde{r}}_A - \tilde{r}_A(abH - b'H\tilde{r}_A + b'\dot{\tilde{r}}_A)}{\tilde{r}_A^2}
\end{equation}
$b = b(\frac{\tilde{r}_A}{a})$. Therefore, for apparent horizon, the variation of total entropy, equation (32), has the following form. 
\vspace{10pt}
\begin{eqnarray}
T_A\dot{S}_{tot} &=& \frac{4\pi \tilde{r}^2_A}{3} \left[  \frac{Hb'\tilde{r}_A^2 - abH\tilde{r}_A - 2H\dot{H}\tilde{r}_A^4}{2H^2\tilde{r}_A^3 - ab + b'\tilde{r}_A}           \right] X(\tilde{r}_A,t)  \ \nonumber \\ \nonumber \\  &-&   \frac{4\pi H\tilde{r}^3_A}{3} Y(\tilde{r}_A,t)
\end{eqnarray}
where 
\begin{eqnarray}
X(\tilde{r}_A,t) &=&    -  6\dot{H}F - 9H^2F + 3H\dot{F} - \frac{3ab'\tilde{r}_AF}{\tilde{r}_A^3}    \ \nonumber \\
&+&  \frac{(ab - \tilde{r}_A)F''}{\tilde{r}_A} +  \frac{(3ab'\tilde{r}_A+3ab-4\tilde{r}_A)F'}{2\tilde{r}_A^2} + \frac{3(\tilde{r}_A-ab)F'^2}{2F\tilde{r}_A}  \ \nonumber \\ &+&   3H^2F'\tilde{r}_A - (\ddot{F})'\tilde{r}_A + \frac{ab''F }{\tilde{r}_A}   + \frac{3F(\dot{F}^2)'\tilde{r}_A - 3F'\dot{F}^2\tilde{r}_A}{2F^2}
\end{eqnarray}
and 
\begin{eqnarray}
Y(\tilde{r}_A,t) &=&    \frac{3(ab-ab'\tilde{r}_A)F}{\tilde{r}_A^3} +  \frac{(\tilde{r}_A - ab)F''}{\tilde{r}_A} +  \frac{(6ab-4\tilde{r}_A)F'}{2\tilde{r}_A^2}   \ \nonumber \\  
&+&   3H^2F'\tilde{r}_A - (\ddot{F})'\tilde{r}_A + \frac{ab''F }{\tilde{r}_A}   + \frac{3F(\dot{F}^2)'\tilde{r}_A - 3F'\dot{F}^2\tilde{r}_A}{2F^2}   \ \nonumber \\  
&-& \frac{6(\tilde{r}_A-ab)F'^2}{2F\tilde{r}_A}
\end{eqnarray}
The expression (33) shows that the generalized second law (GSL) of thermodynamics is fulfilled
in Palatini formalism of modified $f(R)$ gravity  if the right hand side (r.h.s.) of the equation (33) is positive.
So, the validity of GSL depends on the r.h.s. of the equation. 
\vspace{20pt}

\subsection{Thermodynamics at Event Horizon}
In this subsection we need to find out the radius of event horizon and then analyze the generalized
second law of thermodynamics in Palatini version of $f(R)$ gravity. We can consider the following
relation in order to find event horizon with radius $\tilde{r}_E$. 
\begin{equation}
\int_0^{\frac{{\tilde{r}}_E}{a}} \ \frac{dr}{\sqrt{1-\frac{b(r)}{r}}}  \  = \ \int_t^\infty \ \frac{dt}{a}
\end{equation}
Using equation (31), for event horizon, one can reach the following expression for the variation of total entropy. 
\begin{eqnarray}
T_E\dot{S}_{tot} &=& \frac{4\pi \tilde{r}^2_E}{3} \left[  \tilde{r}_E H  -  \sqrt{ 1 - \frac{ab(\frac{\tilde{r}_E}{a})}{\tilde{r}_E} }         \right] X(\tilde{r}_E,t)  \ \nonumber \\ \nonumber \\  &-&   \frac{4\pi H\tilde{r}^3_E}{3} Y(\tilde{r}_E,t)
\end{eqnarray} 
Where $X(\tilde{r}_E,t)$ and $Y(\tilde{r}_E,t)$ are equations (34) and (35), respectively, with the change of $\tilde{r}_A \rightarrow \tilde{r}_E$.
From the above equation, it is clearly seen that the generalizes second law of thermodynamics will be valid
for event horizon in Palatini  $f(R)$ theory  of gravity if the right hand side of the equation (37) is non-negative. 
\vspace{30pt}

\section{Conclusion}
In this work we investigated the thermodynamic properties of time-dependent wormhole geometries
in Palatini version of modified $f(R)$ gravity theories. We derived the gravitational field equations
and then applied to wormhole spacetimes. After that, the thermodynamic behavior of these kinds of phenomena has been
discussed. An expression for the variation of total entropy was found that can express the validity of the
generalized second law (GSL) of thermodynamics. This expression was applied to apparent and event horizons to
check the validity of GSL. For these horizons, it is shown that generalized second law will be fulfilled if
the right hand side of the equations (33) and (37) are non-negative. Investigating the thermodynamic behavior of
wormhole spacetimes in various gravitational theories will be interesting in order to fully understand the nature
of these sorts of theories.
\vspace{30pt}

\textbf{References}
\vspace{20pt}

{01} M. S. Morris and K. S. Thorne, Am. J. Phys. \textbf{56}, 395 (1988).

{02} M. Visser, {\it ``Lorentzian Wormholes"}, Springer-Verlag (1996).

{03} S. V. Sushkov,  \emph{Phys. Rev. D} \textbf{71},  043520 (2005).

{04} F. S. N. Lobo,  \emph{Phys. Rev. D} \textbf{71}, 084011 (2005).

{05} B. Bhawal  and S. Kar,  {\it Phys. Rev.} D {\bf 46}  2464 (1992).

{06} K. A. Bronnikov and  S. W. Kim, {\it Phys. Rev.} D {\bf 67}  0640027 (2003).

{07} F. S. N. Lobo arXiv:1112.6333 [gr-qc].

{08} L. A. Anchordoqui, D. F. Torres, M. L. Trobo and S. E.P. Bergliaffa, {\it Phys. Rev.} D {\bf 57}  829 (1998).

{09} M. Cataldo, P. Salgado and P. Minning, {\it Phys. Rev.} D \textbf{66}, 124008 (2002).

{10} B. Bahawal and S. Kar, {\it Phys. Rev.} D \textbf{46}, 2464-2468 (1992).

{11} N. Godani and G. C.Samanta, {\it New Astronomy, }  \textbf{80}, 101399 (2020).

{12} M. Sharif and S. Rani {\it Mod. Phys. Lett A} \textbf{29}(27)  1450137 (2014).

{13} G. Mustafa, M. R. Shahzad, G. Abbas and T. Xia, {\it Mod. Phys. Lett A }  \textbf{35}(07)  2050035 (2020).

{14} M. Zubair, Rabia Saleem, Yasir Ahmad and G. Abbas  {\it International Journal of Geometric Methods in Modern Physics}, \textbf{16}(03)  1950046 (2019).

{15} C. Bambi, A.  Cardenas-Avendano, G. J. Olmo, D. Rubiera-Garcia {\it Phys. Rev.} D \textbf{93}, 064016 (2016). arXiv:1511.03755 [gr-qc]

{16} H. Saiedi and B.  Nasr Esfahani,  {\it Mod. Phys. Lett A} \textbf{26}(16)  1211 (2011).

{17}  R. Bousso, {\it Phys. Rev.} D \textbf{71}, 064024 (2005).

{18} Y. S. Piao, {\it Phys. Rev.} D \textbf{74}, 047301 (2006).

{19} D. N. Vollick, {\it Phys. Rev.} D \textbf{76}, 124001 (2007).

{20} B. Guberina, R. Horvat and H. Nikolic, {\it Phys.Lett.} B \textbf{636}, 80 (2006).

{21} B. Wang, Y. Gong and E. Abdalla, {\it Phys. Rev.} D \textbf{74}, 083520 (2006).

{22} R. Brustein, {\it Phys. Rev. Lett.} \textbf{84}, 2072 (2000).

{23} H. M. Sadjadi, {\it Phys. Rev.} D \textbf{73}, 063525(2006).

{24} J. Dutta, S. Mitra and B.  Chetry, {\it International Journal of Theoretical Physics},  \textbf{55}, 4272–4285(2016)

{25} J. D. Bekenstein, {\it Phys. Rev.} D \textbf{9}, 3292-3300 (1974).

{26} W. Unruh and R. M. Wald, {\it Phys. Rev.} D \textbf{25}, 942-958 (1982).

{27} J. D. Barrow, {\it Nucl. Phys.} B \textbf{310}, 743-763 (1988).

{28} P. C. W. Davies, {\it Class. Quantum Grav.} \textbf{5}, 1349 (1988); \textbf{4}, L255 (1987).

{29} K. Bamba, C.-Q. Geng, S. Nojiri and S.D. Odintsov, {\it E. P. L.} \textbf{89}(5), 50003 (2010).

{30} A. Paranjape, S. Sarkar and T. Padmanabhan, {\it Phys. Rev.} D \textbf{74}, 104015 (2006).

{31} H. Saiedi, {\it Mod. Phys. Lett A} \textbf{27}(38)  1250220 (2012).

{32} Y. Zheng and R. Yang, {\it Eur.  Phys.  J. C}, \textbf{78} 682 (2018).

{33} T. Bandyopadhyay, U. Debnath, M. Jamil, F. Rahman and R. Myrzakulov, {\it Int. J. Theor. Phys.} \textbf{54} 1750 (2015).

{34} M. Zubair and A. Jawad, {\it Astrophys Space Sci}, \textbf{360} 11 (2015).

{35}  M. Cataldo et al., {\it Phys. Rev.}  D \textbf{79}, 024005 (2009).

{36}  G. J. Olmo, \emph{Int. J. Mod. Phys. D} \textbf{20}, 413 (2011).

{37}  S. Capozziello, F.Darabi, D.Vernieri,  \emph{Mod.Phys.Lett.A} \textbf{25}, 3279-3289 (2010).

{38} H. Saiedi, {\it JHEPGC}, \textbf{6} No. 4 (2020).

{39} S. A. Hayward, {\it Class. Quant. Grav.} \textbf{15}, 3147 (1998).

{40} G. Izquierdo and D. Pavon, \emph{Phys. Lett. B} \textbf{633}, 420 (2006).

\end{document}